\documentclass[aps,pra,showpacs,superscriptaddress,preprint]{revtex4}

\usepackage{amsmath}

\usepackage{subfig}

\usepackage{graphicx}

\catcode`ð=\active
\defð{\u{g}}
\catcode`Ð=\active
\defÐ{\u{G}}
\catcode`Ý=\active
\defÝ{\. I}
\catcode`ö=\active
\defö{\"{o}}
\catcode`Ö=\active
\defÖ{\"O}
\catcode`ü=\active
\defü{\"{u}}
\catcode`Ü=\active
\defÜ{\"{U}}
\catcode`Þ=\active
\defÞ{\c{S}}
\catcode`þ=\active
\defþ{\c{s}}
\catcode`ý=\active
\defý{{\i}}
\catcode`ç=\active
\defç{\d{c}}
\catcode`Ç=\active
\defÇ{\d{C}}

\begin{document}
\title{Scattering of Woods-Saxon Potential in Schrödinger Equation}
\author{\small Altuð Arda}
\email[E-mail: ]{arda@hacettepe.edu.tr}\affiliation{Department of
Physics Education, Hacettepe University, 06800, Ankara,Turkey}
\author{\small Oktay Aydoðdu}
\email[E-mail: ]{oktaydogdu@gmail.com}\affiliation{Department of
Physics, Mersin University, 33343, Mersin,Turkey}
\author{\small Ramazan Sever}
\email[E-mail: ]{sever@metu.edu.tr}\affiliation{Department of
Physics, Middle East Technical  University, 06531, Ankara,Turkey}

\begin{abstract}
The scattering solutions of the one-dimensional Schrödinger
equation for the Woods-Saxon potential are obtained within the
position-dependent mass formalism. The wave functions,
transmission and reflection coefficients are calculated in terms
of Heun's function. These results are also studied for the
constant mass case in detail.
\end{abstract}

\maketitle
\newpage

\section{Introduction}
The quantum mechanical systems could be investigated in the view
of two basic points. One of them is the studying of bound states
to handle the necessary information about the system under
consideration. The other point is solving of scattering problem
for a given quantum mechanical system under the effect of a
potential. So, one has to study both of bound states and
scattering states of a quantum mechanical system under
consideration to achieve a complete information about it. Some
efforts have been made about the scattering problem for a
relativistic and/or non-relativistic system under the influence of
different types of potentials, such as Manning-Rosen potential [1,
2], Eckart potential [3, 4], Pöschl-Teller potential [5],
Hulth\'{e}n potential [6], Woods-Saxon potential [7-9], cusp
potential [10], and Coulomb potential [11]. The scattering problem
in the case where the mass depends on spatially coordinate has
become a particular part of that problem, and has been received
great attention to study of scattering states for a given quantum
system [12-15]. The position-dependent mass formalism is a useful
ground to explain the electronic properties of quantum wells and
quantum dots [16], semiconductor heterostructures [17], and
impurities in crystals [18-20].

In this paper, we solve the following one-dimensional Schrödinger
equation ($\hbar=1$)
\begin{eqnarray}
\bigg\{\,\frac{d^2}{dx^2}-\frac{dm(x)/dx}{m(x)}\,\frac{d}{dx}+2m(x)[E-V(x)]\bigg\}\psi(x)=0\,,
\end{eqnarray}
obtained from the Hamiltonian [14]
\begin{eqnarray}
H=\frac{1}{2}\bigg(\hat{p}\,\frac{1}{m}\,\hat{p}\bigg)+V\,.
\end{eqnarray}
for the Woods-Saxon potential to study the scattering states
within the framework of position-dependent mass formalism. The
effective-mass Schrödinger equation could be transformed into
Heun's equation [21] which is a Fuchsian-type equation with four
singularities [14] by using a coordinate transformation. We obtain
the wave function in terms of Heun's function and then we find
transmission and reflection coefficients by studying the
asymptotic behavior of the wave function at infinity. We write
also the transmission and reflection coefficients for the case of
constant mass by using the properties of Heun's function and also
the continuity conditions of the wave function at $x=0$. We find
the wave function for the case of constant mass in terms of
hypergeometric functions and plot the wave functions for
completeness. In nuclear physics, the Woods-Saxon potential is
used to construct a shell model to describe the single-particle
motion in a fusing system [22] and the potential plays an
important role within the microscopic physics because of
describing the interaction of a nucleon with a heavy nucleus [23].

The work is organized as follows. In Section II we obtain exactly
scattering state solutions of the Woods-Saxon potential and
transmission and reflection coefficients in the case of
position-dependent mass. We study also the same quantities in the
case of constant mass. The conclusions are given in Section III.
In Appendix A we list some equalities related with Heun's function
required for this work.
\section{Scattering State Solutions}
The Woods-Saxon potential has the form
\begin{eqnarray}
V(x)=-\frac{V_{0}}{1+e\,^{\delta x}}\,,
\end{eqnarray}
and we parameterize the mass function as
\begin{eqnarray}
m(x)=(m_{0}-m_{1})\big(M-\frac{1}{1+e\,^{\delta x}}\big)\,,
\end{eqnarray}
where $M=(m_{0}+m_{1})/(m_{0}-m_{1})$ and $V_{0},\delta,m_{0}$ and
$m_{1}$ are positive parameters. The form of the mass function is
strongly similar to that of the potential. We could exactly solve
the problem because of this form and also study the results for
the case of constant mass. By using the transformation
$y=(1+e\,^{\delta x})^{-1}$ and inserting Eq.(4) and Eq.(3) into
Eq.(1), we obtain the differential equation ($0<y<1$)
\begin{eqnarray}
&&\psi''(y)+\bigg(\frac{1}{y}+\frac{1}{y-1}-\frac{1}{y-M}\bigg)\psi'(y)\nonumber\\&+&
\frac{1}{y(y-1)(y-M)}\big\{-a^2_{1}y-\frac{a^2_{2}}{y}M+\frac{a^2_{3}}{y-1}\,(M-1)+Ma^2_{1}+a^2_{2}-a^2_{3}
\big\}\psi(y)=0\,,
\end{eqnarray}
where
\begin{eqnarray}
a^2_{1}=(2/\delta^2)(m_{0}-m_{1})V_{0}\,\,;\,\,-a^2_{2}=(2/\delta^2)(m_{0}+m_{1})E\,\,;\,\,
-a^2_{3}=(4/\delta^2)m_{1}(E+V_{0})\,.
\end{eqnarray}
To obtain a Fuchsian-type differential equation from Eq. (5), we
use a new transformation
\begin{eqnarray}
\psi(y)=y^{a_{2}}(y-1)^{a_{3}}f(y)\,,
\end{eqnarray}
which gives a Heun's-type equation given as in Eq. (A1) in
Appendix A
\begin{eqnarray}
&&f''(y)+\bigg(\frac{1+2a_{2}}{y}+\frac{1+2a_{3}}{y-1}-\frac{1}{y-M}\bigg)f'(y)\nonumber\\&+&
\frac{1}{y(y-1)(y-M)}\bigg\{[-a^2_{1}+(a_{2}+a_{3})^2]y-[-a^2_{1}+(a_{2}+a_{3})(1+a_{2}
+a_{3})]M+a_{2}\bigg\}f(y)=0\,.\nonumber\\
\end{eqnarray}
The general solution of Eq. (8), which is regular in the
neighborhood of $y=0$, is written in terms of the Heun's function
as [14]
\begin{eqnarray}
f(y)=AH(M,-[-a^2_{1}&+&(a_{2}+a_{3})(1+a_{2}+a_{3})]M+a_{2};a_{2}+a_{3}-a_{1},\nonumber\\
&&a_{2}+a_{3}+a_{1},1+2a_{2},-1;y)\,,
\end{eqnarray}
where the constant $A$ will be determined below.

Let us first investigate the limit $x \rightarrow \infty
\big(y\simeq e^{-\delta x}\rightarrow 0\big)$, which gives
$f(0)=A$ in Eq. (9) and the solution $\psi(y) \rightarrow A
y\,^{a_{2}}=Ae^{-\delta a_{2}x}$ becomes
\begin{eqnarray}
\psi(x)=Ae^{-ik_{1}x}\,,
\end{eqnarray}
where $k_{1}=\sqrt{2(m_{0}+m_{1})E\,}$ and we have used the
property of $H(a,b;\alpha,\beta,\gamma,\delta;0)=1$.

To study the behavior of the solution Eq. (9) for $x \rightarrow
-\infty (y \rightarrow 1), 1-y\simeq e^{\delta x}$, we use Eq.
(A5) of Appendix A, which changes the argument $y$ to $1-y$. Thus,
we obtain the Heun's function in Eq. (9) as
\begin{eqnarray}
&&H(M,-[-a^2_{1}+(a_{2}+a_{3})(1+a_{2}+a_{3})]M+a_{2};a_{2}+a_{3}-a_{1},a_{2}+a_{3}+a_{1},1+2a_{2},-1;y)
=\nonumber\\&&D_{1}H(1-M,[-a^2_{1}+(a_{2}+a_{3})(1+a_{2}+a_{3})]M+a^2_{1}-(a_{2}+a_{3})^2-a_{2};\nonumber\\
&&a_{2}+a_{3}-a_{1},a_{2}+a_{3}+a_{1},1+2a_{3},-1;1-y)+D_{2}(1-y)^{-2a_{3}}\nonumber\\&&\times
H(1-M,[-a^2_{1}+(a_{2}-a_{3})(1+a_{2}-a_{3})]M+a^2_{1}-(a_{2}-a_{3})^2-a_{2};\nonumber\\
&&a_{2}-a_{3}+a_{1},a_{2}-a_{3}-a_{1},1-2a_{3},-1;1-y)\,,
\end{eqnarray}
where the constants $D_{1}$ and $D_{2}$ are written by using Eq.
(A6) in Appendix A
\begin{eqnarray}
D_1&=&H(M,-[-a^2_{1}+(a_{2}+a_{3})(1+a_{2}+a_{3})]M+a_{2};\nonumber\\&&a_{2}+a_{3}-a_{1},
a_{2}+a_{3}+a_{1},1+2a_{2},-1;1)\,,\\
D_2&=&H(M,-[-a^2_{1}+(a_{2}-a_{3})(1+a_{2}-a_{3})]M+a_{2};\nonumber\\&&a_{2}-a_{3}+a_{1},
a_{2}-a_{3}-a_{1},1+2a_{2},-1;1)\,.
\end{eqnarray}

Using Eq. (11) we obtain the solution in Eq. (9) as
\begin{eqnarray}
\psi(y) \rightarrow A(-1)^{a_{3}}\{D_{1}e^{\delta
a_{3}x}+D_{2}e^{-\delta a_{3}x}\}\,,
\end{eqnarray}
which gives
\begin{eqnarray}
\psi(x)=e^{ik_{2}x}+\frac{D_2}{D_1}\,e^{-ik_{2}x}\,,
\end{eqnarray}
where $k_{2}=\sqrt{4m_{1}(E+V_{0})\,}$  and we set
$A=(-1)^{-a_{3}}/D_{1}$. Thus, we achieve the following form of
the wave function for the limit $x \rightarrow \pm\infty$
\begin{eqnarray}
\psi(x)=\left\{
\begin{array}{lr}
e^{ik_{2}x}+Re^{-ik_{2}x}\,, & x \rightarrow -\infty\,,\\
T'e^{ik_{1}x}\,, & x \rightarrow +\infty\,,\\
\end{array}\right.
\end{eqnarray}
As a result, we recover the asymptotic behavior of a plane wave
coming from the left-hand side.

We can write the wave function explicitly
\begin{eqnarray}
\psi(x)&=&(-1)^{2a_{3}}(1+e^{\delta x})^{-(a_{2}+a_{3})}e^{\delta
a_{3}x}\nonumber\\&\times&\frac{H(M,b+a_{2};a_{2}+a_{3}-a_{1},a_{2}+a_{3}+a_{1},1+2a_{2},-1;\frac{1}{1+e^{\delta
x}})}{H(M,b+a_{2};a_{2}+a_{3}-a_{1},a_{2}+a_{3}+a_{1},1+2a_{2},-1;1)}\,,
\end{eqnarray}
where $b=-[-a^2_{1}+(a_{2}+a_{3})(1+a_{2}+a_{3})]M$. Finally, we
give the reflection and transmission coefficients for the case of
position-dependent mass, respectively
\begin{eqnarray}
|R|^2=\left|\frac{H(M,b'+a_{2};a_{2}-a_{3}+a_{1},a_{2}-a_{3}-a_{1},1+2a_{2},-1;1)}{H(M,b+a_{2};a_{2}+a_{3}-a_{1},
a_{2}+a_{3}+a_{1},1+2a_{2},-1;1)}\right|^2\,,
\end{eqnarray}
where $b'=-[-a^2_{1}+(a_{2}-a_{3})(1+a_{2}-a_{3})]M$, and
\begin{eqnarray}
|T|^2=\frac{k_1}{k_2}\,\frac{1}{\left|H(M,b+a_{2};a_{2}+a_{3}-a_{1},a_{2}+a_{3}+a_{1},1+2a_{2},-1;1)\right|^2}\,.
\end{eqnarray}

In order to investigate the dependence of the reflection
coefficient to the energy $E$, we rewrite Eq. (18) in the
following form by interchanging $\alpha \leftrightarrow \beta$ in
Heun's function
\begin{eqnarray}
|R|^2&=&\left|\frac{H(M,b'+a_{2};a_{2}-a_{3}+a_{1},a_{2}-a_{3}-a_{1},1+2a_{2},-1;1)}{H(M,b+a_{2};a_{2}+a_{3}+a_{1},
a_{2}+a_{3}-a_{1},1+2a_{2},-1;1)}\right|\nonumber\\&\times&
\left|\frac{H(M,b'+a_{2};a_{2}-a_{3}-a_{1},a_{2}-a_{3}+a_{1},1+2a_{2},-1;1)}{H(M,b+a_{2};a_{2}+a_{3}-a_{1},
a_{2}+a_{3}+a_{1},1+2a_{2},-1;1)}\right|\,,
\end{eqnarray}
By using Eq. (A7) in Appendix A and keeping in mind that
$a^2_{1}=a^2_{2}/M-a^2_{3}/(M-1)$, Eq. (20) gives
\begin{eqnarray}
|R|^2&=&\frac{[(M-1)a_{2}+Ma_{3}]^2}{[(M-1)a_{2}-Ma_{3}]^2}\nonumber\\
&\times&\left|\frac{H(M,b'+a_{3}-a_{1};a_{2}-a_{3}+a_{1},a_{2}-a_{3}-a_{1}+1,2+2a_{2},0;1)}{H(M,b-a_{3}-a_{1};a_{2}+a_{3}
+a_{1},
a_{2}+a_{3}-a_{1}+1,2+2a_{2},0;1)}\right|\nonumber\\&\times&
\left|\frac{H(M,b'+a_{3}+a_{1};a_{2}-a_{3}-a_{1},a_{2}-a_{3}+a_{1}+1,2+2a_{2},0;1)}{H(M,b-a_{3}+a_{1};a_{2}+a_{3}-a_{1},
a_{2}+a_{3}+a_{1}+1,2+2a_{2},0;1)}\right|\,,
\end{eqnarray}
This equation enables us to analyze the dependence of reflection
coefficient to the energy $E$ when the energy goes to infinity. In
this case, Eq. (21) gives
\begin{eqnarray}
|R|^2&=&_{E \rightarrow
\infty}\bigg(\frac{\sqrt{2m_{1}\,}-\sqrt{m_{0}+m_{1}\,}}{\sqrt{2m_{1}\,}+\sqrt{m_{0}+m_{1}\,}}\bigg)^2
\left|\frac{H(M,-M(a_{2}-a_{3})^2;a_{2}-a_{3},a_{2}-a_{3},2a_{2},0;1)}{H(M,-M(a_{2}+a_{3})^2;a_{2}+a_{3},
a_{2}+a_{3},2a_{2},0;1)}\right|^2\,,\nonumber\\
\end{eqnarray}
Using the equality
$H(a,b;\alpha,\beta,\gamma,0;y)=\,_2F_{1}(\alpha,\beta;\gamma;y)$
(for $b=-a\alpha\beta$) [14] and also
$_2F_{1}(\alpha,\alpha;\gamma;1)\rightarrow_{\alpha,\gamma
\rightarrow \infty}e\,^{\alpha^2/\gamma}$, we obtain
\begin{eqnarray}
|R|^2&=&_{E \rightarrow
\infty}\bigg(\frac{\sqrt{2m_{1}\,}-\sqrt{m_{0}+m_{1}\,}}{\sqrt{2m_{1}\,}+\sqrt{m_{0}+m_{1}\,}}\bigg)^2\,.
\end{eqnarray}
Eq. (23) shows that the reflection coefficient increases up to the
value obtained in Eq. (23) while changing with energy. Fig. (1)
shows the variation of the reflection and transmission coefficients
as a function of the energy $E$ in the position dependent mass case.
In Fig. (2), the effect of the mass parameters $m_0$ and $m_1$ on
the reflection and transmission coefficients are given. It is seen
that the reflection coefficient decreases linearly with mass
parameters while the transmission coefficient increases with the
growing values of the parameters. In the Figs. (1) and (2), It could
be seen that the unitarity condition $|R|^2+|T|^2=1$ is satisfied in
the constant and position dependent nass cases. In Fig. (2), we see
that the reflection coefficient can not take zero value for the case
of $E<V_0$ which is agreed with quantum mechanical results.

Now, we begin to give the results for the case of constant mass,
which means that $m_{0}=m_{1}$, starting from the wave function.
With the help of Eq. (A8) in Appendix A, we write the wave
function
\begin{eqnarray}
\psi(x)_{m_{0}=m_{1}}&=&(-1)^{2a_{3}}(1+e^{\delta
x})^{-(a_{2}+a_{3})}e^{\delta
a_{3}x}\nonumber\\&\times&\frac{_2F_1(1+a_{2}+a_{3},a_{2}+a_{3};1+2a_{2};\frac{1}{1+e^{\delta
x}})}{_2F_1(1+a_{2}+a_{3},a_{2}+a_{3};1+2a_{2};1)}\,,
\end{eqnarray}
which can be written in terms of Gamma functions
\begin{eqnarray}
\psi(x)_{m_{0}=m_{1}}&=&(-1)^{2a_{3}}(1+e^{\delta
x})^{-(a_{2}+a_{3})}e^{\delta
a_{3}x}\frac{\Gamma(a_{2}-a_{3})\Gamma(1+a_{2}-a_{3})}{\Gamma(1+2a_{2})\Gamma(-2a_{3})}
\nonumber\\&\times&_2F_1(1+a_{2}+a_{3},a_{2}+a_{3};1+2a_{2};\frac{1}{1+e^{\delta
x}})\,.
\end{eqnarray}
where used the relation of hypergeometric function
$_2F_1(\alpha,\beta;\gamma;1)=\frac{\Gamma(\gamma)\Gamma(\gamma-\alpha-\beta)}{\Gamma(\gamma-\alpha)
\Gamma(\gamma-\beta)}$. The parameters given in Eq. (6) in the
case of constant mass become ($m_{0}=m_{1}=m$)
\begin{eqnarray}
-a^2_{2}=(4/\delta^2)mE\,\,;\,\,-a^2_{3}=(4/\delta^2)m(E+V_{0})\,.
\end{eqnarray}

We depict the wave function for two different values of parameter
sets in Fig. (3). It is seen that the wave function exhibit an
oscillatory behaviour for $x<0$ and exponentially decreasing in
the region $x>0$. The oscillating behaviour of the wave function
given in Eq. (16) for $x<0$ is a purely quantum mechanical
interference effect between the incident and reflected waves [23].
The wave function in the region $x>0$ goes to zero due to the
potential given in Eq. (3).

We give the reflection and transmission coefficients for the case
of constant mass. Using Eq. (A8) in Appendix A and the relation
$_2F_1(\alpha,\beta;\gamma;1)=\frac{\Gamma(\gamma)\Gamma(\gamma-\alpha-\beta)}
{\Gamma(\gamma-\alpha)\Gamma(\gamma-\beta)}$ in Eq. (18), we
obtain
\begin{eqnarray}
\left|R\right|^2_{m_{0}=m_{1}}=\left|\frac{_2F_1(a_{2}-a_{3}+1,a_{2}-a_{3};1+2a_{2};1)}
{_2F_1(a_{2}+a_{3}+1,a_{2}+a_{3};1+2a_{2};1)}\right|^2
=\left|\frac{\Gamma(2a_{3})\Gamma(a_{2}-a_{3})\Gamma(a_{2}-a_{3}+1)}{\Gamma(-2a_{3})\Gamma(a_{2}+a_{3})
\Gamma(a_{2}+a_{3}+1)}\right|^2\,,\nonumber\\
\end{eqnarray}
and similarly from Eq. (19)
\begin{eqnarray}
\left|T\right|^2_{m_{0}=m_{1}}=\frac{k_1}{k_2}\,\frac{1}{\left|\,_2F_1(a_{2}+a_{3}+1,a_{2}+a_{3};1+2a_{2};1)\right|^2}
=\frac{k_1}{k_2}\,\left|\frac{\Gamma(a_{2}-a_{3})\Gamma(a_{2}-a_{3}+1)}{\Gamma(1+2a_{2})\Gamma(-2a_{3})}
\right|^2\,.\nonumber\\
\end{eqnarray}

It should be noted that we must apply the continuity condition to
obtain a relation between the coefficients written in Eq. (16). The
condition that the wave function and its derivative must be
continuous at $x=0$ gives $k_{2}(1-|R|^2)=k_{1}|T'|^2$ [24, 25]. In
Fig. (4), we plot the variation of the reflection and transmission
coefficients according to the energy $E$ in the case of constant
mass. The reflection coefficient goes to zero when the energy
increases while the transmission coefficient goes to unity. It could
be interesting to study the limiting case of $\delta \rightarrow
\infty$. In that case the potential function becomes $V(x)
\rightarrow 0$ and the mass function goes to $2m$. It means that the
reflection and transmission can not appear (Eqs. (27) and (28)) as
expected. In addition, in the limiting case $\delta \rightarrow
-\infty$ we obtain a step potential from Eq. (3) and Eq. (4) gives
us $m(x) \rightarrow 2m$. Thus, we get the reflection coefficient as

\begin{eqnarray}
\left|R\right|^2_{m_{0}=m_{1}}=_{\delta \rightarrow
-\infty}\left|\frac{a_{2}-a_{3}}{a_{2}+a_{3}}\right|^2=_{\delta
\rightarrow
-\infty}\left(\frac{k_{1}-k_{2}}{k_{1}+k_{2}}\right)^2\,.
\end{eqnarray}
where $k_1=\sqrt{4mE}$ and $k_2=\sqrt{4m(E+V_0)}$.

\section{Conclusion}

We have exactly solved the one-dimensional effective mass
Schrödinger equation for the Woods-Saxon potential. We have found
the wave functions in terms of Heun's function. The reflection and
transmission coefficients are calculated by using the asymptotic
behaviour of the wave function at infinity. To analyze these
coefficients in the case of position-dependent mass, we calculate
the reflection coefficient in the limit $E \rightarrow \infty$. They
are plotted as a function of mass parameters in Fig. (2). One can
see that the unitarity condition in the scattering problem given as
$|R|^2+|T|^2=1$ is satisfied in the position dependent mass case
also. We have also obtained the wave function, reflection and
transmission coefficients in the constant mass case. They are
presented in the Figs. (3) and (4).

\section{Acknowledgments}
This research was partially supported by the Scientific and
Technical Research Council of Turkey. The authors would like to
thank the referee whose comments help us to improve this work.

\appendix
\section{Useful Equalities of Heun's Function}
Heun's equation with the following form
\begin{eqnarray}
\bigg\{\,\frac{d^2}{dy^2}+\bigg(\,\frac{\gamma}{y}+\frac{1+\alpha+\beta-\gamma-\delta}{y-1}-\frac{\delta}{y-a}\bigg)
\frac{d}{dy}+\frac{\alpha\beta y+b}{y(y-1)(y-a)}\bigg\}f(y)=0\,,
\end{eqnarray}
has a solution in the neighborhood of $y=0$
\begin{eqnarray}
f(y)=H(a,b;\alpha,\beta,\gamma,\delta;y)\,,
\end{eqnarray}
and two linearly independent solutions in the neighborhood of
$y=1$ [14]
\begin{eqnarray}
f(y)=H(1-a,-b-\alpha\beta;\alpha,\beta,1+\alpha+\beta-\gamma-\delta,\delta;1-y)\,,
\end{eqnarray}
and
\begin{eqnarray}
f(y)=(1-y)^{\gamma+\delta-\alpha-\beta}H(1-a,-b-\alpha\beta-(\gamma+\delta-\alpha-\beta)
(\gamma+\delta-a\gamma);\gamma+\delta-\alpha,\nonumber\\\gamma+\delta-\beta,1-\alpha-\beta+\gamma+\delta,\delta;1-y)\,,
\end{eqnarray}

The solution in the neighborhood of $y=0$ can be written as a
linear combination of last two Heun's functions [14]
\begin{eqnarray}
H(a,b;\alpha,\beta,\gamma,\delta;y)&=&D_{1}H(1-a,-b-\alpha\beta;\alpha,\beta,1+\alpha+\beta-\gamma-\delta,\delta;1-y)
\nonumber\\
&+&D_{2}(1-y)^{\gamma+\delta-\alpha-\beta}\nonumber\\&\times&H(1-a,-b-\alpha\beta-(\gamma+\delta-\alpha-\beta)
(\gamma+\delta-a\gamma);\nonumber\\&&\gamma+\delta-\alpha,\gamma+\delta-\beta,1-\alpha-\beta+\gamma+\delta,\delta;1-y)\,,
\end{eqnarray}
where the constants are given
\begin{eqnarray}
D_1&=&H(a,b;\alpha,\beta,\gamma,\delta;1)\,,\nonumber\\
D_2&=&H(a,b-a\gamma(\gamma+\delta-\alpha-\beta);\gamma+\delta-\alpha,\gamma+\delta-\beta,\gamma,\delta;1)\,.
\end{eqnarray}

The following identity links the arguments $(\beta, \gamma,
\delta)$ to $(\beta+1, \gamma+1, \delta+1)$, respectively,
\begin{eqnarray}
(\gamma
a\beta+b)H(a,b-\alpha;\alpha,\beta+1,\gamma+1,\delta+1;y)\nonumber\\=
a\gamma
H(a,b;\alpha,\beta,\gamma,\delta;y)+a\gamma(y-1)\frac{d}{dy}H(a,b;\alpha,\beta,\gamma,\delta;y)\,.
\end{eqnarray}

Finally, in the limit of $a \rightarrow \infty$, Heun's function
turns into a hypergeometric function [14]
\begin{eqnarray}
H(a,a\Delta;\alpha,\beta,\gamma,\delta;y)=_{a \rightarrow \infty}
\,_2F_{1}\bigg(\,\frac{1}{2}(\alpha+\beta-\delta)+\sqrt{[\frac{1}{2}(\alpha+\beta-\delta)]^2+\Delta\,}\,,\nonumber\\
\frac{1}{2}(\alpha+\beta-\delta)-\sqrt{[\frac{1}{2}(\alpha+\beta-\delta)]^2+\Delta\,};\gamma;y\bigg)\,.
\end{eqnarray}
with $\gamma\neq-n(n=0,1,2,...)$.

\newpage

\begin{figure}[htbp]
\centering
\includegraphics[height=3.5in, width=6in, angle=0]{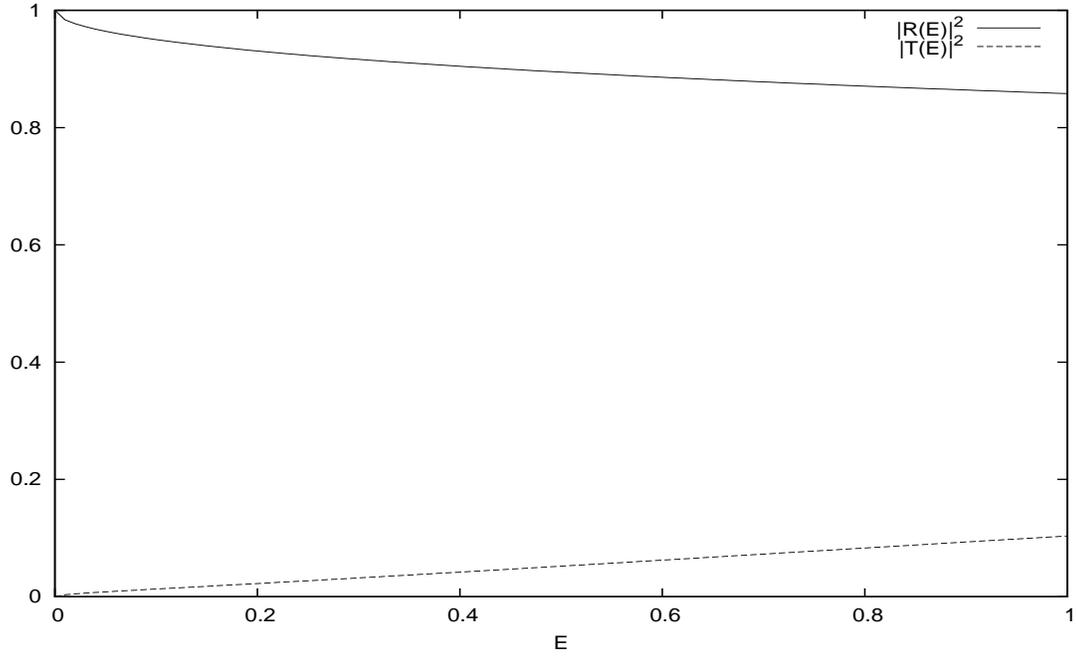}
\caption{The reflection and transmission coefficients in the case
of position-dependent mass for $m_0=0.1, m_1=10, \delta=5,
V_{0}=5$.}
\end{figure}

\newpage

\begin{figure}
\centering \subfloat[][reflection and transmission coefficients
for $m_1=0.01, \delta=3, V_{0}=1$ and $E=0.05$.]{\includegraphics[height=3.5in, width=6in, angle=0]{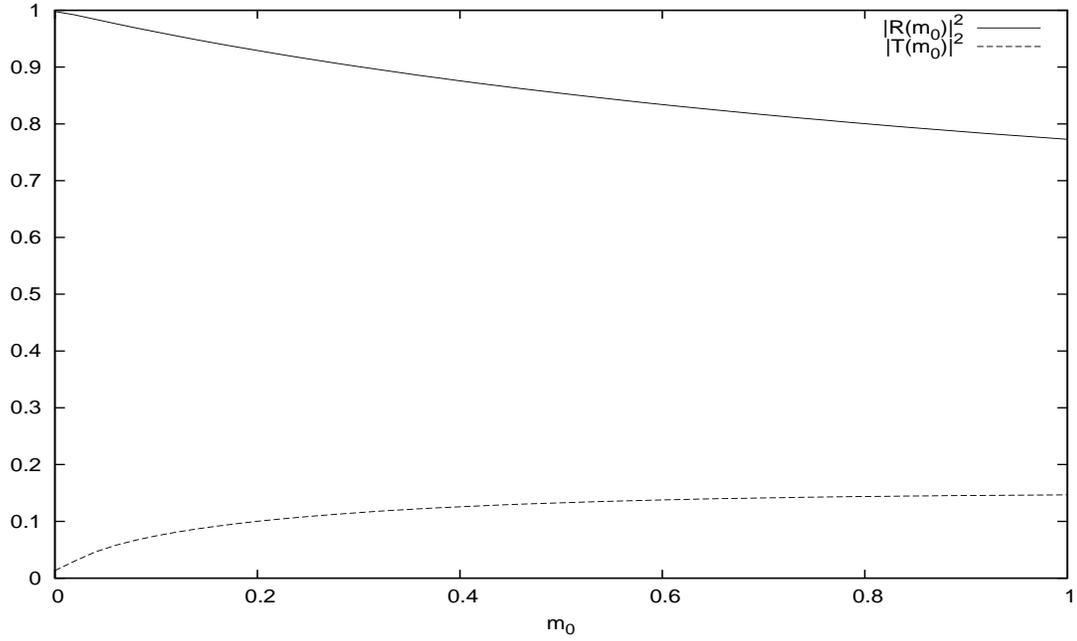}}\\
\subfloat[][reflection and transmission coefficients for $m_0=1,
\delta=5, V_{0}=1$ and $E=0.1$ .]{\includegraphics[height=3.5in,
width=6in, angle=0]{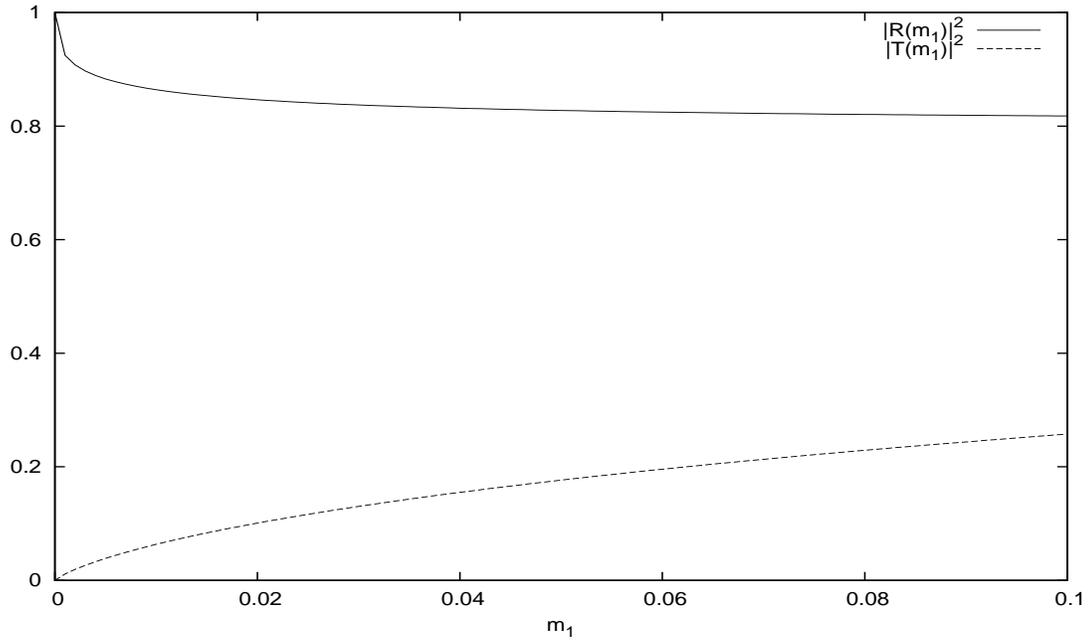}} \caption{Reflection and transmission
coefficients in the case of position-dependent mass.}
\end{figure}

\begin{figure}[htbp]
\centering
\includegraphics[height=3.5in, width=6in, angle=0]{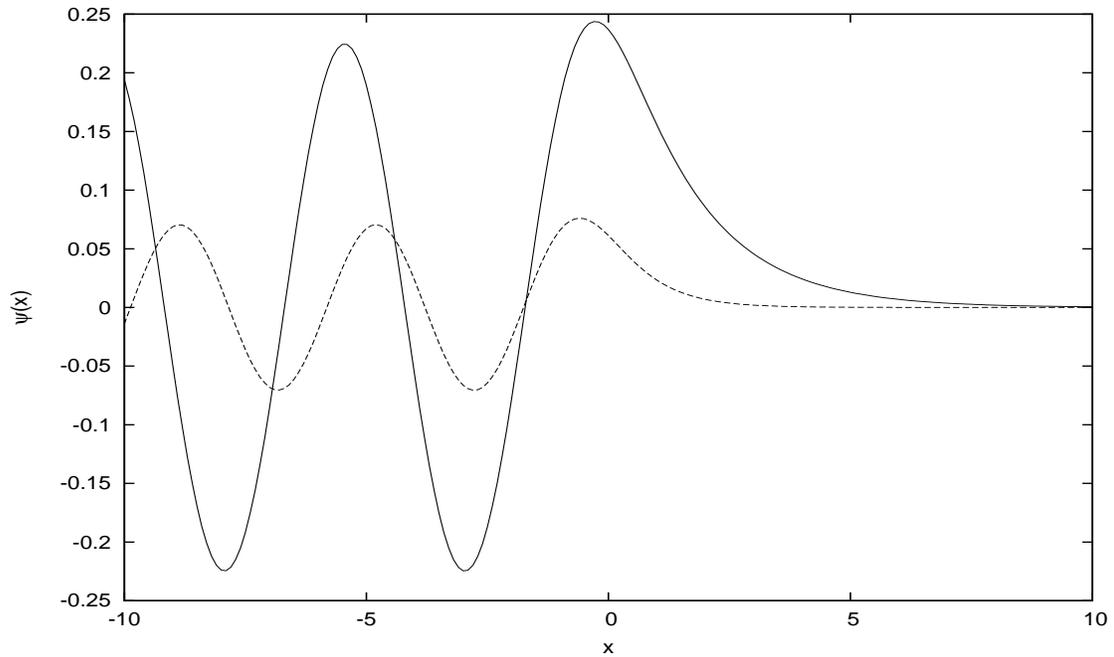}
\caption{The unnormalized wave functions in the case of constant
mass for $m=1, \delta=2, V_{0}=0.5, E=-m/10$ (solid line) and for
$m=2, \delta=2, V_{0}=0.5, E=-m/10$.}
\end{figure}

\newpage

\begin{figure}
\centering \subfloat[][reflection and transmission coefficients
for $m=0.5, \delta=5, V_{0}=1$.]{\includegraphics[height=3.5in,
width=6in, angle=0]{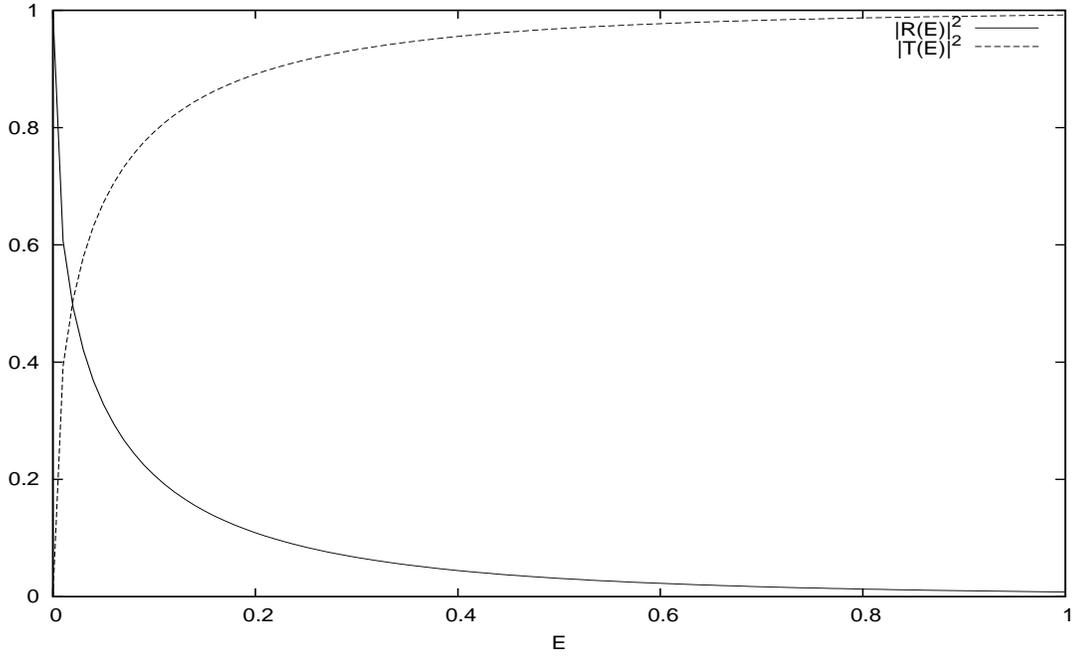}}\\
\subfloat[][reflection and transmission coefficients for $m=1,
\delta=5, V_{0}=1$.]{\includegraphics[height=3in, width=6in,
angle=0]{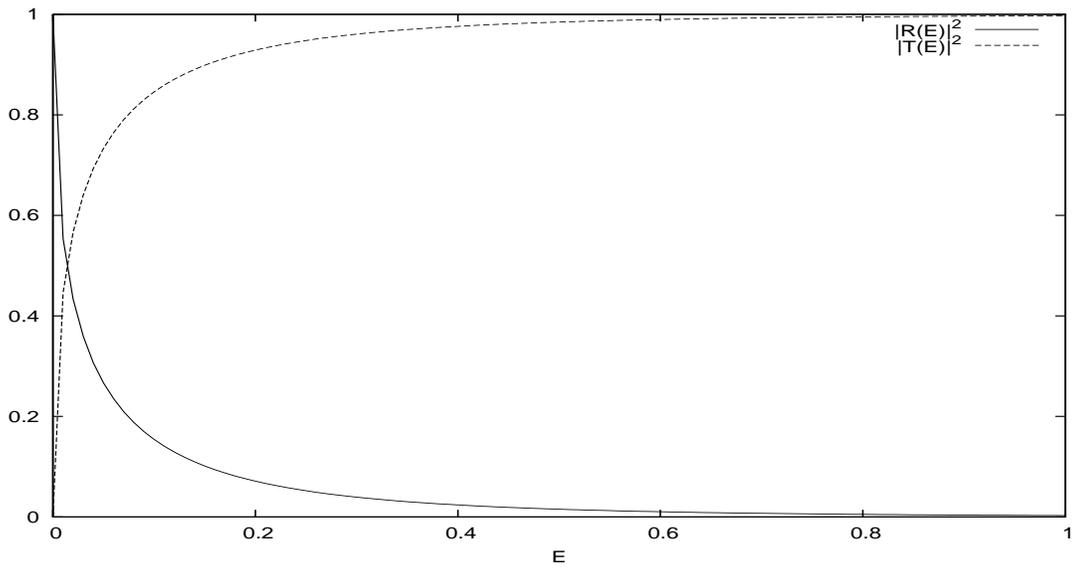}} \caption{Reflection and transmission coefficients
in the case of constant mass.}
\end{figure}

\end{document}